# Amplitude- and phase-resolved nano-imaging and nano-spectroscopy of polaritons in liquid environment

Divya Virmani[1], Andrei Bylinkin[1], Irene Dolado Lopez[1], Eli Janzen[2], James H. Edgar[2], and Rainer Hillenbrand[3,4]

[1]CIC nanoGUNE BRTA, Tolosa Hiribidea 76, 20018 Donostia-San Sebastián, Spain.

[2] Kansas State University, Tim Taylor Department of Chemical Engineering, Durland Hall, Manhattan, KS 66506 USA

[3]CIC nanoGUNE BRTA and Department of Electricity and Electronics, UPV/EHU, Tolosa Hiribidea 76, 20018 Donostia-San Sebastián, Spain.

[4]IKERBASQUE, Basque Foundation for Science, 48011 Bilbao, Spain.

**Abstract**: Localized and propagating polaritons allow for highly sensitive analysis of (bio)chemical substances and processes. Nanoimaging of the polaritons' evanescent fields allows for critically important experimental mode identification and for studying field confinement. Here we describe two setups for polariton nanoimaging and spectroscopy in liquid, which is an indispensable environment for (bio)chemical samples. We first demonstrate antenna mapping with a transflection infrared scattering-type scanning near-field optical microscope (s-SNOM), where the tip acts as a near-field scattering probe. We then demonstrate a total internal reflection (TIR) based setup, where the tip is both launching and probing ultra-confined polaritons in van der Waals materials, here phonon polaritons in hexagonal boron nitride (h-BN) flakes. This work lays the foundation for s-SNOM based polariton interferometry in liquid, which has wide application potential for in-situ studies of chemical reactions at the bare or functionalized surface of polaritonic materials, including (bio)chemical recognition analogous to the classical surface plasmon resonance spectroscopy.

**Keywords**: scattering type scanning near-field optical microcopy (s-SNOM), nano-FTIR, liquid environment, polaritons, plasmonic antennas

Surface plasmon resonance (SPR) spectroscopy using metal surfaces is a widely used label-free technique for studying biomolecular interactions in their native environment, that is, in liquids[1–4]. Typically, SPR spectroscopy is performed in the visible spectral range. On the other hand, localized plasmon resonances in metal antennas have recently also boosted surface-enhanced infrared absorption spectroscopy (SEIRA) not only in air[5,6] but also in liquids[7–9], allowing for highly sensitive analysis of chemical composition and molecular conformations of minute amounts of a substance, offering new possibilities for plasmonic sensing in the mid-infrared (mid-IR) spectral range. More recently, ultra-confined plasmon and phonon polaritons in van der Waals (vdW) materials such as graphene, hexagonal boron nitride (h-BN) and molybdenum trioxide ($MoO_3$) have attracted wide attention[10–14], as they promise ultra-compact and highly sensitive devices for field-enhanced spectroscopy, molecular sensing and signal processing at infrared frequencies[15–18]. The majority of these studies were performed in air. Considering the tremendous impact of SPR sensing in liquids, it would be of great interest to explore the properties of ultra-confined vdW material polaritons in liquid environment, for future exploitation, for example, in mid-IR biosensing applications.

Graphene plasmons and phonon polaritons in polar vdW materials can be imaged and studied in great detail by scanning probe techniques, such as scattering-type scanning near-field optical microscopy (s-SNOM), photothermal expansion (PTE) microscopy and photo-induced force microscopy (PiFM). So far, polariton imaging in the aqueous phase has been demonstrated only by PTE[19] and PiFM[20] and considered not being possible by s-SNOM. On the other hand, the interferometric detection scheme of s-SNOM offers many advantages over PTE and PiFM, including direct near-field phase measurements for retrieving phase and group velocities of polaritons[21], reconstruction of complex-valued dielectric functions[22,23], and tomographic sample reconstruction[24]. Further, only s-SNOM has been demonstrated so far for nanoimaging and nanospectroscopy in the broad spectral range from visible to sub-THz frequencies, utilizing radiation from continuous wave (CW), ultrafast lasers[25,26], and synchrotrons[27,28], or free electron lasers[29].

For these reasons, various efforts have been made to develop techniques that enable s-SNOM to study samples in liquid environment. Mid-IR s-SNOM imaging and spectroscopy of objects immersed in liquids has been first achieved with the help of graphene-based liquid cells, where the s-SNOM operation was performed as usual in air[30–32]. More recent developments demonstrated a more general operation principle, where both the near-field probe as well as samples are immersed in liquid[33,34]. Such implementations – following the basic concepts of well-established atomic force microscopy in liquid – in the future may allow for versatile imaging of even larger objects (such as whole cells) and avoid potential near-field interaction between the samples in the liquid and the membranes covering them. However, the reported techniques are not well established yet, and challenged by various technical and scientific aspects, for example. complicated beam shapes were required when illuminating under normal incidence allowing nanoimaging but not spectroscopy[34], while TIR-based geometry was successfully used for nanospectroscopy but deemed challenging for nanoimaging due to effective beam path change during scanning[33]. Further, polariton imaging has not been demonstrated so far.

Here, we introduce a normal-incidence mid-IR s-SNOM setup – operating with standard mid-IR Gaussian beams and the near-field probe being integrated into the liquid cell – for imaging the near-field distribution of plasmonic metal antennas in liquid environment. We also describe our total internal reflection setup, which allows for efficient coupling of standard mid-IR Gaussian beams to the near-field probe, in order to perform both tip-enhanced infrared nanoimaging and nanospectroscopy. We apply this setup for amplitude- and phase-resolved imaging of ultra-confined propagating phonon polaritons (PhPs) on h-BN flakes, yielding not only the polariton dispersion but also the sign of the phase velocity, the latter not being directly accessible by PTE microscopy.

S-SNOM[35] is an atomic force microscopy (AFM) based technique where a light beam is focused onto the apex of the metallized AFM probe. The tip acts as an antenna and concentrates the radiation to a highly confined and enhanced near-field spot at the tip apex. The near fields interact with the sample and modify the backscattered field depending on the sample's local dielectric properties. The backscattered field is recorded with a pseudo-heterodyne Michelson interferometer[36] as a function of tip position yielding near-field amplitude ($s$) and phase ($\phi$) images simultaneous to the sample topography. Background contributions are suppressed by higher harmonic demodulation of the backscattered field at $n\Omega$ for $n \geq 2$ where, $\Omega$ is the tip oscillation frequency. In a typical s-SNOM experiment[37], the tip is illuminated in a side illumination scheme. However, such an illumination geometry is not suitable for operation in liquid and at infrared frequencies, owing to strong absorption and distortion of the infrared beam by the liquid (e.g. by $H_2O$).

We first describe our transflection s-SNOM setup for infrared antenna mapping (Figure 1a), which is based on a modified commercial s-SNOM (neaSNOM from Neaspec GmbH) setup. We illuminate a standard Pt-Ir coated AFM tip (NCPt arrow tip, Nanoworld) with a parabolic mirror ($f = 8 \text{ mm}$) through the substrate and sample holder ($CaF_2$) at normal incidence with IR light from a tunable $CO_2$ laser (Soliton GmbH). Normal incidence is naturally an ideal geometry for antenna mapping, as it allows for efficient excitation of antennas while minimizing tip excitation (field polarization is perpendicular to the tip axis and thus the tip is not excited efficiently), thus minimizing distortion of antenna fields caused by the tip[38]. This has been previously demonstrated to be ideally suited for infrared antenna mapping in air[39]. The difference in our experimental setup is that the antenna sample is now placed in water, and for that reason we detect in backward direction (i.e. through the substrate, in order to minimize IR beam absorption and distortion in water above the samples). The backscattered light is recorded with a standard pseudo-heterodyne detection scheme employing an MCT detector (InfraRed Associates, Inc.) combined with higher harmonic demodulation (as described earlier). To operate the setup in water, we use an open liquid cell concept, which is simply a droplet of water confined between the sample and a transparent window above the AFM cantilever (that allows for measuring cantilever deflection as part of AFM operation).

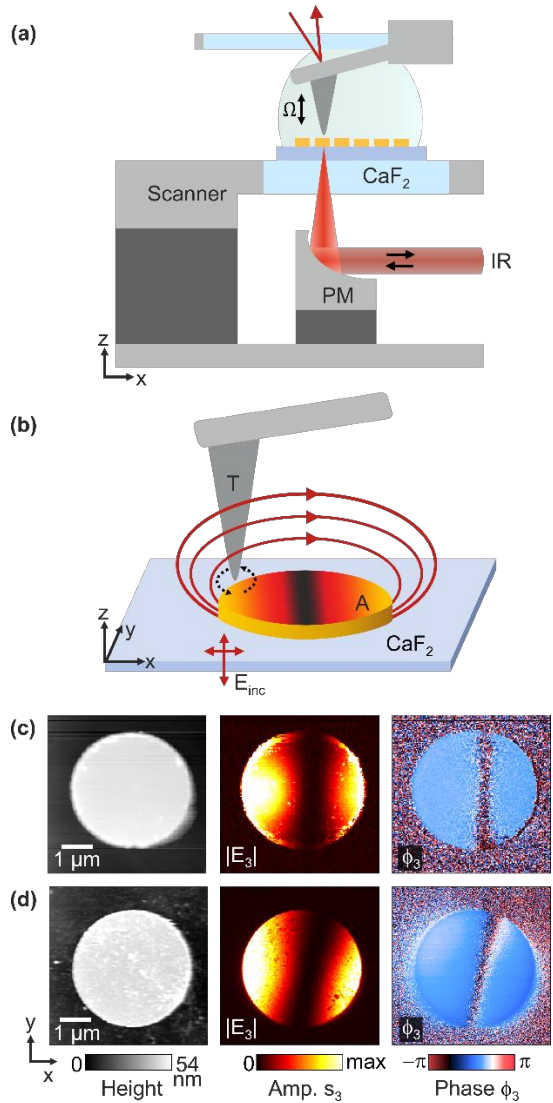

**Figure 1. Normal incidence tansflection s-SNOM for mapping mid-IR plasmonic antennas.** a) Schematic of the setup. PM, parabolic mirror; $\Omega$, tip oscillation frequency. b) Illustration of near-field scattering process. The incident field $E_{inc}$ illuminates the antenna (A). The antenna fields illuminate and polarizes the tip (T), which scatters the near fields via the antenna into the far field where it is detected interferometrically (not shown here). c,d) Topography, amplitude and phase (from left to right, respectively) images of the dipolar antenna mode of a gold disk (c) in $H_2O$ and (d) in air. Mid-IR imaging wavelength is 10.5 µm.

We demonstrate the mapping of infrared antenna fields in liquid with a gold disk of 4 µm diameter on $CaF_2$, which acts as a dipole-resonant antenna at 10.5 µm wavelength. Simultaneously to the clear topography image (Figure 1c, left), we observe two bright spots in the amplitude image (Figure 1c, middle). They are aligned along the polarization of the incident beam (indicated by double sided arrow in Figure 1b) and reveal the strongly enhanced fields of the in-plane oriented dipole. The phase (Figure 1c, right) at the position of the two bright spots is stable and nearly the same. Similar amplitude and phase images were obtained for the same disk imaged in air (Figure 1d), demonstrating that our setup allows for reliable amplitude- and phase-resolved imaging of infrared antenna modes without any distortion by the liquid (apart from potential small frequency shifts due to the liquid environment, which do not affect the mode pattern). Comparing with previous antenna mapping s-SNOM experiments[40,41], we find that the amplitude and phase pattern resemble the ones reported for side-illumination s-SNOM operating with s-polarized light (where particularly the phase is the same for the two bright amplitude spots). This can be explained by a double scattering process[40,41], which is illustrated in Figure 1b. The near field of the antenna illuminates the tip and polarizes it. This additional polarization can be approximated as an antenna-induced dipole in the tip. Due to its close proximity to the antenna, it radiates into the far field (and thus into the far-field detector of our setup) via the gold disk. Thus, the gold disk acts twice, first enhancing the tip-illumination and then enhancing the tip-scattering. For that reason, the amplitude image yields essentially the sum of square of the in- and out-of-plane electric antenna near fields (weighted by the polarizability tensor of the tip) and a constant phase across the whole disk. A more detailed discussion goes beyond the scope of this paper, and we refer the interested reader to a detailed analysis provided in Ref. 40.

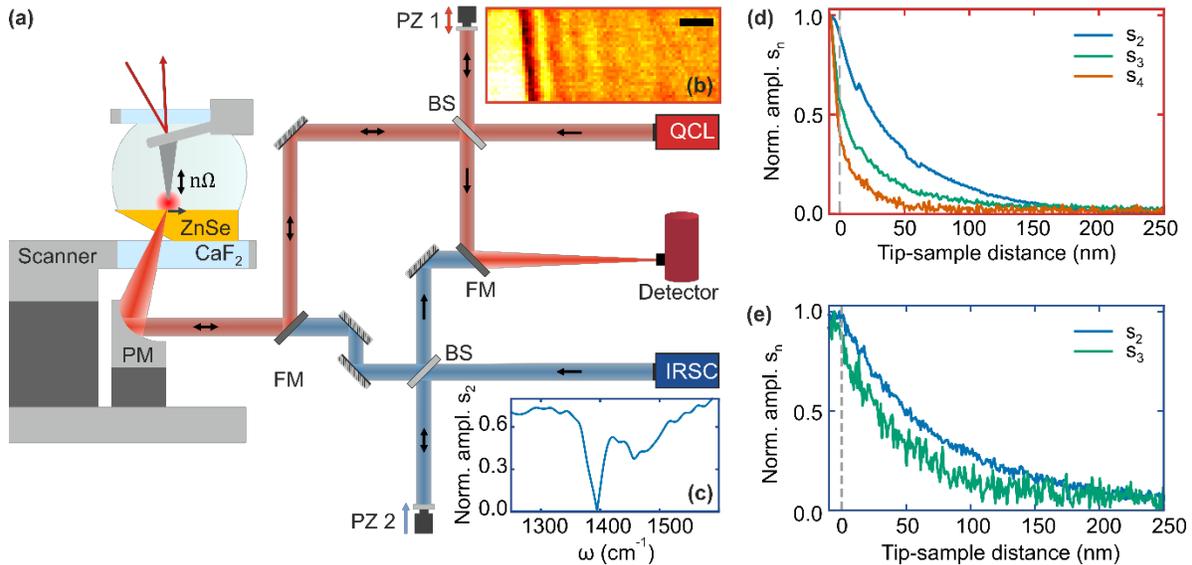

**Figure 2. TIR based s-SNOM and nano-FTIR setup for imaging and spectroscopy in liquid** a) Schematic of the setup for nanoimaging-(s-SNOM) and nanospectroscopy (nano-FTIR). For s-SNOM, the tip is illuminated with a frequency-tunable QCL. The backscattered light is demodulated at tip frequency $n\Omega$ and recorded with a pseudo-heterodyne Michelson interferometer (top). BS, beamsplitter; PZ1, piezo actuated vibrating mirror. For nano-FTIR, the tip is illuminated with an infrared supercontinuum (IRSC) broadband laser. The backscattered light is analyzed using an asymmetric Fourier transform spectrometer (bottom). PZ2, piezo actuated translating mirror; FM, flip mirror. b) Example of an s-SNOM amplitude image $s_3$. Scalebar: $500\ nm$. c) Example of a nano-FTIR amplitude spectrum. d) Approach curves showing the decay of s-SNOM amplitude signals $s_n$ in liquid with increasing tip sample distance recorded at $1600\ cm^{-1}$. Tapping amplitude (TA)~$60\ nm$. e) Approach curves for nano-FTIR signal (at white light position) in liquid. TA ~$120\ nm$.

The normal incidence transflection mode s-SNOM works efficiently for amplitude- and phase-resolved nanoimaging of the near-field distribution of resonant plasmonic structures, where the near fields are generated by the antenna and the tip has the exclusive role of being a scattering probe. However, for imaging ultra-confined polaritons in 2D materials, where the tips often also play the important role of being the polariton launcher, an efficient coupling of the incident light with the tip is necessary to provide a strongly concentrated field at the tip apex. In normal incidence illumination, such coupling is weak, as the polarization of the incident field is perpendicular to the tip axis. It turns out that an efficient tip excitation can be achieved by illuminating via total internal reflection (TIR), as previously described by Ref. 33 for spectroscopic chemical imaging of biological samples. In the following, we demonstrate that this scheme can well be applied for both amplitude- and phase-resolved nanoimaging and nanospectroscopy of ultra-confined phonon polaritons (PhPs), specifically hyperbolic phonon polaritons in thin layers of h-BN.

A schematic of our TIR based setup is shown in Figure 2a. An off-axis parabola ($f = 10\ \text{mm}$) is used to focus the light into a 2 mm thick ZnSe ($n = 2.4$) wedge at an angle of 50° relative to the tip axis. The ZnSe wedge was produced by polishing one of the faces of a ZnSe substrate at an angle of 30°, in order to illuminate ZnSe/D$_2$O interface at an angle $\theta > \theta_c \approx 34°$ relative to the tip axis. This was done mostly due to space constraints in the current setup. The sample of interest can then be placed directly onto the wedge, which subsequently is placed on a CaF$_2$ window and the same liquid cell concept as described in Figure 1a is applied. The tip is illuminated via the evanescent fields generated by the total internally reflected beam at the sample-liquid interface. The backscattered beam is collected via the same parabolic mirror and directed to the detector to form either a nanoscale resolved image at a single frequency (nanoimaging), or a nanoscale resolved Fourier transform infrared (nano-FTIR) spectra of the sample.

For infrared nanoimaging (Figure 2a, top), the tip is illuminated with a monochromatic IR light of a frequency tunable quantum cascade laser (QCL, Daylight Solutions Inc.), the backscattered field is recorded in a similar fashion to our normal incidence transflection setup yielding near-field amplitude (shown in Figure 2b) and phase images, $s_n$ and $\phi_n$, respectively. Note that the wedged ZnSe substrate with the sample is raster scanned during the imaging process causing an effective beam path change in the y direction due to the varying thickness of the substrate that the beam transverses (see geometry illustrations in Figures 3a and 3b). This leads to a phase gradient superimposed on the phase images, which has been subtracted from all the phase images shown in this work (Supplementary Figure S1). For nano-FTIR spectroscopy (Figure 2a, bottom), the tip is illuminated with a broadband infrared radiation from a difference frequency generated laser supercontinuum (frequency range: $1200 - 1700\ cm^{-1}$, average power: $500\mu W$). The backscattered light is recorded with an asymmetric Fourier transform spectrometer, yielding amplitude (shown in Figure 2c) and phase spectra, $s_n(\omega)$ and $\phi_n(\omega)$, after normalization to corresponding reference spectra recorded on the bare substrate (ZnSe).

Figures 2d and 2e display typical approach curves recorded with the s-SNOM and nano-FTIR setup, respectively, showing the demodulated near-field amplitude signal $s_n$ ($n \geq 2$) as a function of the tip-sample distance. A strong signal decay with increasing tip-sample distance can be seen, which is stronger for higher harmonic signals, analogous to what is well known from s-SNOM in air[42]. At large tip-sample distances, all curves decay towards the noise floor, verifying that background-free near-field signals are measured. The more rapid signal decay at higher harmonic signal demodulation in air is well known to significantly improve the surface sensitivity and spatial resolution of s-SNOM and nano-FTIR. Figures 2d and 2e thus show that higher harmonic demodulation may be exploited for increasing the spatial resolution and surface sensitivity in liquid as well.

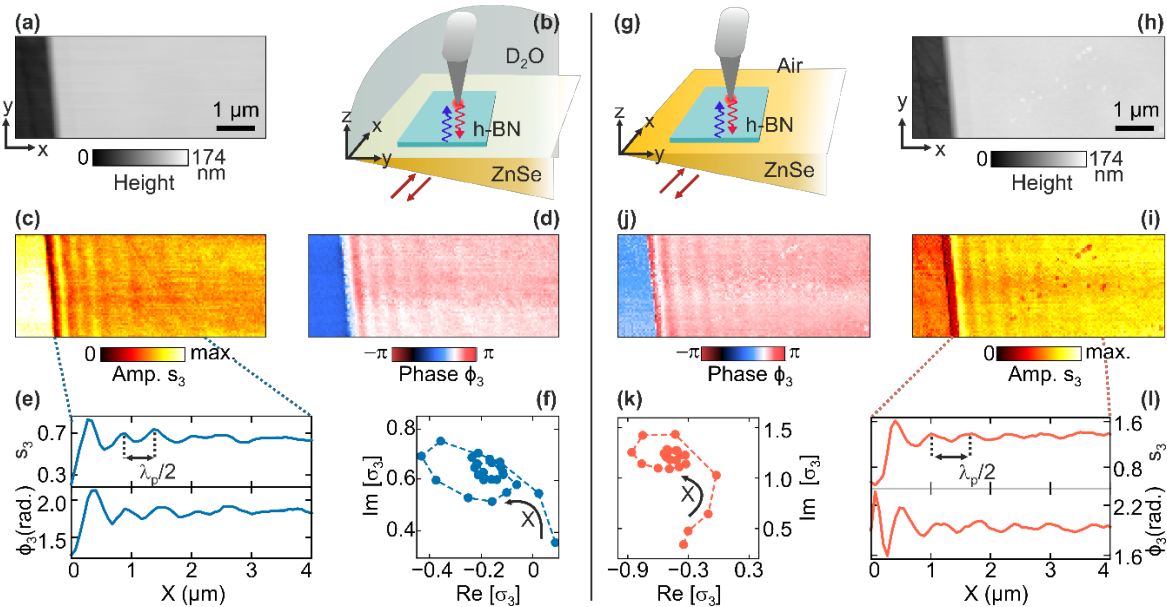

**Figure 3. Amplitude- and phase-resolved imaging of h-BN PhPs in liquid (a-f) and air (g-l).** a) AFM topography of a 110 nm thick h-BN flake recorded in D$_2$O. b) Schematics of probing in liquid. c,d) Near-field amplitude and phase images recorded at $1486\ cm^{-1}$ in D$_2$O, revealing a standing wave pattern caused by the PhP interference. e) Amplitude and phase line profiles extracted perpendicular to the h-BN edge (averaged over 50 lines), revealing a fringe spacing of $\lambda_p/2$. e) Phasor diagram of the complex-valued near-field signal $\sigma_3 = s_3\ e^{i\phi_3}$, indicating a positive phase velocity of h-BN PhPs launched in D$_2$O. g-l) Same as a-f, but probing was done in air.

Propagating hyperbolic phonon polaritons (PhPs) in liquid were imaged with exfoliated h-BN flakes, which were transferred directly onto the top surface of the ZnSe wedge. The topography image in Figure 3a clearly shows the smooth surface of the 110 nm thick h-BN flake with little noise, indicating a stable tapping-mode AFM operation in liquid. In the simultaneously recorded amplitude $s_3$ and phase $\phi_3$ images at $1486 \text{ cm}^{-1}$ (Figures 3c,d) we observe fringes on the h-BN surface, which are parallel to the h-BN edge, similar to s-SNOM observations of h-BN PhPs in air[12,43]. The fringes can be explained by the interference of PhPs, as illustrated in Figure 3b. Typically, the interference of edge and tip launched PhPs form standing wave patterns, leading to a fringe spacing of either $\lambda_P/2$ or $\lambda_P$, respectively, where $\lambda_P$ is the wavelength of the phonon polaritons. Importantly, and in contrast to PTE measurements of phonon polaritons in liquid, amplitude and phase mapping enabled by s-SNOM allows us to directly determine the phase velocity of PhPs in liquid. To that end, we extract amplitude and phase line profiles (Figure 3e) from the images and plot them in the complex plane (phasor diagram, Figure 3f). An anticlockwise rotation is observed, corresponding to positive phase velocity with increasing the tip-edge distance x. For comparison, we imaged the same flake at the same IR frequency in air. The images look similar, but the fringe spacing in the infrared images recorded in air is slightly larger than in the images recorded in liquid. This is expected, as D$_2$O has a higher permittivity compared to air, thus acting as a dielectric load[19]. Dielectric materials are well known and understood to reduce the polariton wavelength with increasing permittivity[44,45].

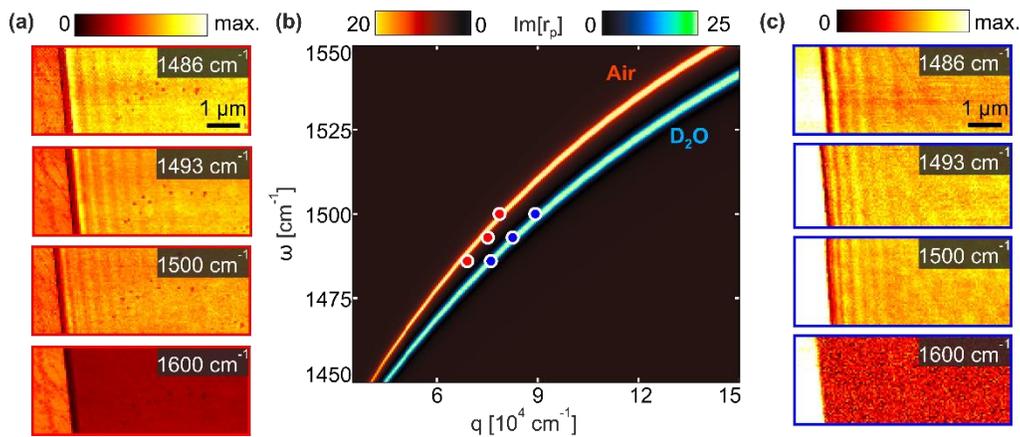

*Figure 4. Dispersion of h-BN PhPs in liquid and air.* *a) Near-field amplitude images $s_3$ of an h-BN flake in air. b) Symbols show PhP momenta obtained from line profiles taken from the images shown in panel a and c. The color plot indicates the calculated imaginary part of the momentum- and frequency-dependent Fresnel reflection coefficient, Im[r$_p$], of a 90 nm thick h-BN flake on ZnSe in air (red) and D$_2$O(blue). c) Near-field amplitude images $s_3$ of an h-BN flake in liquid.*

For a quantitative analysis and comparison of the phonon polariton properties in liquid and air, we imaged the h-BN flake at different frequencies (Figure 4a,c shows amplitude images, corresponding phase images are shown in Supplementary Figure S2).To measure the polariton momentum $q = 2\pi/\lambda_P$ at each frequency, we extracted line profiles perpendicular to the h-BN edge and Fourier transformed them. The corresponding PhP momenta are plotted in Figure 4b (symbols), clearly showing the systematic shift of the experimental dispersion to higher wavevectors when the h-BN is imaged in liquid. Note that fringes are absent in the images at $1600 \text{ cm}^{-1}$, as this frequency is close to the longitudinal phonon frequency of h-BN, where ultra-confined PhPs are too confined to be observed. We confirm our observations by calculating (see Methods) the dispersion of the fundamental PhP mode (in literature often denoted M0) of an h-BN flake on ZnSe in either D$_2$O or air (blue and red curves in Figure 4b). The experimental data (symbols) and the calculated dispersion (lines) are in good agreement with each other for an h-BN thickness of 90 nm, which is smaller than the experimental h-BN thickness, which we attribute to uncertainties in height measurements and dielectric data for the permittivities of h-BN and ZnSe.

To demonstrate nano-FTIR spectroscopic imaging in liquid, we recorded a line scan perpendicular to the edge of the h-BN flake. Figure 5 shows the normalized nano-FTIR amplitude spectra as a function of tip position x. We clearly see a horizontal bright feature centered at $1395\ cm^{-1}$, which corresponds to the h-BN phonon resonance. More interestingly, we observe fringes, which we attribute to a PhP interference pattern, as the fringe spacing $\lambda_p/2$ decreases continuously with increasing frequency. We find a good qualitative agreement to similar datasets recorded in air[12]. Figure 5 clearly demonstrates that hyperspectral nano-FTIR imaging of ultra-confined polaritons in liquid is possible. However, we refrain form a quantitative analysis of this data set, as the required long data acquisition time (1 hour and 20 minutes) and significant sample heating led to strong sample drift during nano-FTIR recording. In future developments and applications, this challenge of sample drift can be tackled by improving thermal stability of our setup and employing drift correction strategies[46]. Furthermore, the heating of the liquid could be reduced by employing heat sinks or implementing flow cells instead of the simple liquid droplet used in the presented work. We finally note that dielectric (materials) contrasts (e.g. between h-BN and ZnSe) obtained with our setup in liquid differ from those in air. Future studies are required to explore this phenomenon, which could be caused by illumination of the tip by both propagating and evanescent fields due to a non-optimal prism geometry (due to space restrictions in our current setup) or enhanced higher mechanical harmonic generation in liquid.

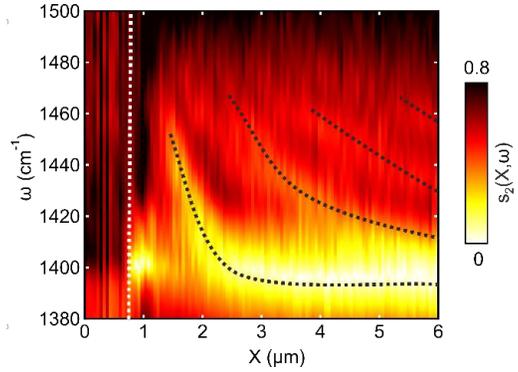

*Figure 5. nano-FTIR spectral line scan of h-BN in liquid.* The nano-FTIR amplitude spectra $s_2(\omega)$ were recorded as a function of distance x between tip and h-BN edge (indicated by the dashed vertical line). Black dashed lines are guides to the eye, tracing the interference fringes. Spectral resolution: $14\ cm^{-1}$; Total accusation time $\sim 1hr\,20min$.

To summarize, we demonstrated two methods enabling infrared s-SNOM imaging of polaritons in liquid. We first demonstrated a normal incidence transflection s-SNOM for background-free amplitude- and phase-resolved infrared nanoimaging of the near-field distribution of resonant plasmonic antennas in liquid. By tuning the plasmonic resonances to match the molecular vibrations, such a system could be used in the future for in-situ s-SNOM studies of (bio)chemical substances with enhanced sensitivity[47], vibrational strong coupling experiments[48], or plasmon-induced chemical modifications. We also demonstrated a TIR-based s-SNOM for nanoimaging and spectroscopy of ultra-confined h-BN phonon polaritons in liquid. We presented how the presence of liquid modifies the PhP dispersion. In future, one could exploit polariton interferometry in liquid (pure solutions or buffer solutions containing specific molecules) for studying chemical interactions at the interface between the bare or functionalized polaritonic materials. Finally, we note that the TIR setup could be further optimized for standard s-SNOM and nano-FTIR based chemical mapping and identification of samples that are fully immersed in liquid.

**Methods**

**Dispersion calculation.** The h-BN dispersion shown in Figure 4b was calculated by quasi-normal mode analysis using the COMSOL mode solver. In the mode analysis we consider infinitely long h-BN layers. From the mode solver we extracted the mode index $\frac{q}{q_0} + i\gamma/q_0$, which is the complex-valued wavevector normalized to the photon momentum $q_0$, which we use to calculate the polariton wavelength $\lambda_p = 2\pi/q$. The dielectric permittivity of isotopically ($^{10}$B) enriched h-BN was calculated by Lorentz model according to $\epsilon_{h-BN,j}(\omega) = \epsilon_{\infty,j}\left(\frac{\omega_{LO,j}^2 - \omega^2 - i\omega\Gamma_j}{\omega_{TO,j}^2 - \omega^2 - i\omega\Gamma_j}\right)$, where $j = \perp, \parallel$ indicates the direction of the tensor with respect to the anisotropy axis. $\omega_{LO,j}$ and $\omega_{TO,j}$ are the LO and TO phonon frequencies, $\Gamma$ is the damping constant and $\epsilon_{\infty,j}$ is the high frequency permittivity. The h-BN parameters used for the calculations are as follows: $\omega_{LO,\perp} = 1650\ cm^{-1}, \omega_{LO,\parallel} = 845\ cm^{-1}, \omega_{TO,\perp} = 1394.5\ cm^{-1}, \omega_{TO,\parallel} = 785\ cm^{-1}, \Gamma_\perp = 1.8\ cm^{-1}, \Gamma_\parallel = 1\ cm^{-1}$, and $\epsilon_{\infty,\parallel} = 2.5$ taken from the Ref. 49. A modified value of $\epsilon_{\infty,\perp} = 4.5$ was used instead of 5.1 in Ref. 49 to better match to our experiments. The permittivity of ZnSe ($\epsilon = 5.76$) and D$_2$O ($\epsilon = 1.8225$) at a wavelength of 6 μm were taken from Ref. 50 and Ref. 51, respectively.

**Sample fabrication. Gold Antennas.** We fabricated Au disks on the CaF$_2$ substrate via high-resolution electron-beam lithography. Polymethyl methacrylate (PMMA) was spin coated onto the substrate at 4000 rpm as the electron-sensitive polymer. The PMMA was subsequently covered by a 2-nm-thick layer of gold for enabling lithography on the insulating substrate. After the electron-beam assisted writing of the disks, the gold was chemically etched (5 s immersion in KI/I2 solution) and the PMMA was developed in methyl isobutyl ketone: isopropanol 1:3. Finally 3 nm layer of Ti were deposited by electron beam evaporation followed by thermal evaporation of 40 nm of Au. The lift-off of the disks was done by immersing the sample in acetone overnight.

**Exfoliation and transfer of h-BN flakes.** We used isotopically ($^{10}$B) enriched h-BN[52]. The h-BN crystals were grown by the atmospheric pressure metal flux growth method as described previously in Ref. 53,54. We performed mechanical exfoliation of the h-BN crystals using blue Nitto tape (Nitto Europe NV, Genk, Belgium). We then performed a second exfoliation of the h-BN flakes from the tape directly onto the ZnSe substrate. Isolated homogeneous h-BN flakes were then used for this work.

## Acknowledgements

This work has received funding from the European Union's Horizon 2020 research and innovation program under the Marie Skłodowska Curie grant agreement No. 721874 (SPM2.0). The authors further acknowledge financial support from the Spanish Ministry of Science, Innovation and Universities (national project RTI2018-094830-B-100 and the project MDM-2016-0618 of the Marie de Maeztu Units of Excellence Program) and the Basque Government (grant No. IT1164-19). Support to investigate h-BN crystal growth from the Office of Naval Research award N00014-20-1-2427 is greatly appreciated.

## Competing interests

R.H. is a co-founder of Neaspec GmbH, a company producing scattering-type scanning near-field optical microscope systems, such as the one used in this study. The remaining authors declare no competing interests.

*Supplementary information*

# Amplitude- and phase- resolved nano-imaging and nano-spectroscopy of polaritons in liquid environment

Virmani et.al.

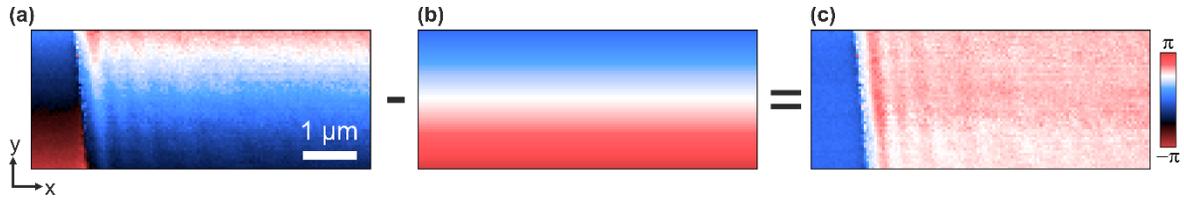

**S 1 Phase drift correction.** *a) Near field phase $\phi_3$ image superimposed by phase gradient caused by the effective beam path change along the y direction during imaging. b) Phase gradient $\sim 60 \times 10^{-3} * y$. c) Corrected phase image obtained after subtracting the gradient in panel b from raw phase image in panel a.*

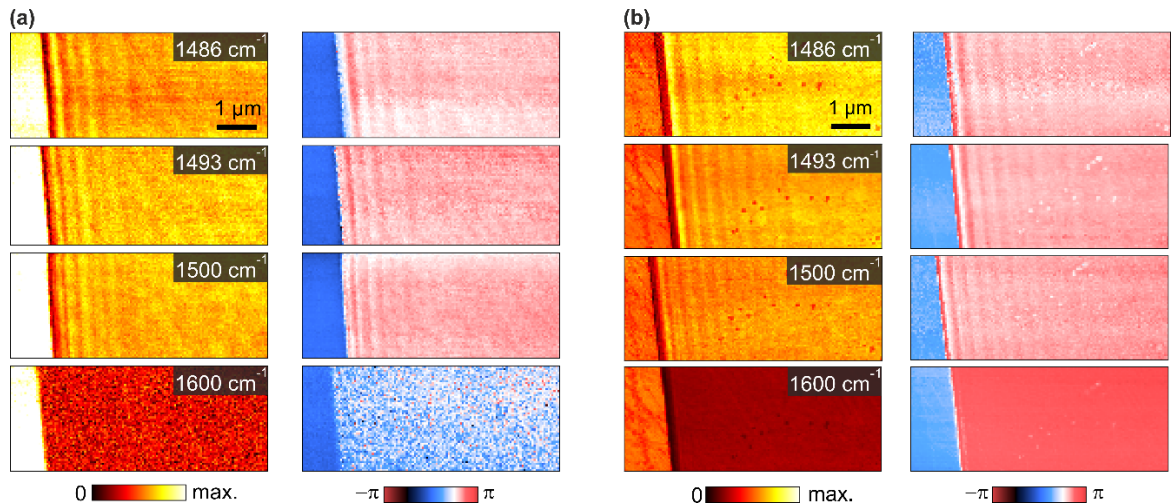

**S 2 s-SNOM nanoimaging in liquid and air.** *a) Near field amplitude (left) and Phase(right) image of h-BN PhPs launched at different frequencies in liquid. b) Images recorded for the same flake at varying frequencies in air.*